\documentclass[twocolumn,amsmath,amssymb,aps,longbibliography,floatfix]{revtex4-1}

\usepackage{cancel}
\usepackage{enumitem}
\usepackage[utf8x]{inputenc}
\usepackage{graphicx}
\usepackage{dcolumn}
\usepackage{bm}
\usepackage{hyperref}
\usepackage[switch]{lineno}
\usepackage{float}             
\usepackage{subfigure}
\usepackage{tikz}

\begin{document}

\title{Ligand-field contributions to spin-phonon coupling in a family of Vanadium molecular qubits from multi-reference electronic structure theory.}

\author{$^{1}$Alessandro Lunghi}
\email{lunghia@tcd.ie}

\affiliation{$^{1}$School of Physics, CRANN Institute and AMBER, Trinity College Dublin, Dublin 2, Ireland}

\begin{abstract}
{\bf Molecular electronic spins represent one of the most promising building blocks for the design of quantum computing architectures. However, the advancement of this technology requires the increase of spin lifetime at ambient temperature. Spin-phonon coupling has been recognized as the key interaction dictating spin relaxation at high temperature in molecular crystals and the search for chemical-design principles to control such interaction are a fundamental challenge in the field. Here we present a multi-reference first-principles analysis of the $g$-tensor and the spin-phonon coupling in a series of four exa-coordinate Vanadium(IV) molecular complexes, where the catecholate ligand donor atom is progressively changed from Oxygen to Sulphur, Selenium and Tellurium. A ligand field interpretation of the multi-reference electronic structure theory results made it possible to rationalize the correlation between the molecular $g$-shifts and the average spin-phonon coupling coefficients, revealing the role of spin-orbit coupling, chemical bond covalency and energy splitting of d-like orbitals in spin relaxation. Our study reveals the simultaneous increase of metal-ligand covalency and electronic excited state energy separation as key elements of an optimal strategy towards long spin-lattice lifetimes in molecular qubits.}
\end{abstract}

\maketitle

\section*{Introduction}

Electronic unpaired spins embedded in solid-state environments are among the most promising building-blocks for the development of quantum-computing architectures thanks to the optimal trade-off between long spin coherence times and the possibility to manipulate them with short external micro-wave pulses\cite{Morton2011,Kloeffel2013}. Coordination complexes have a fundamental strength with respect to the many solid-state platforms that have been proposed to implement qubits and quantum gates: intrinsically-long spin coherence times and the possibility to exploit chemical synthesis in order to further tune their properties or couple them\cite{Ardavan2007,Lehmann2009,Wedge2012,Aromi2012}. Coordination compounds with a spin-$1/2$ ground state therefore represent a natural class of molecular qubits and have extensively been used as a playground to extent spin coherence times by ligand design\cite{Bader2014,Zadrozny2015,Bader2016,Atzori2016b,Atzori2018a,Fataftah2019a}.

Recently, an increasing amount of experimental evidence is pointing to the role of spin-lattice interaction as the main limitation to long coherence times at high-temperature\cite{Bader2014,Bader2016,Atzori2016b,Fataftah2019a}. In particular, it is now clear that the microscopic details of spin-lattice relaxation, due to spin-phonon interaction, are of key importance in the determination of the high-temperature behaviour of spin dynamics and therefore deserve a careful consideration\cite{Escalera-moreno2018}. In this respect, a first-principles spin-phonon relaxation formalism has recently been developed\cite{Escalera-Moreno2017,Lunghi2017,Lunghi2019b}. This computational approach to spin relaxation is paving the way to a quantitative description of spin dynamics in general solid-state spin systems and has already provided fundamental new insights into the intra-molecular nature of spin-phonon coupling and the role of molecular vibrations\cite{Escalera-Moreno2017,Lunghi2017a,Goodwin2017,Lunghi2019b,Escalera-moreno2019}. Although the theoretical treatment of spin-lattice relaxation it is rapidly evolving, a clear understanding of spin-phonon coupling in terms of an intuitive chemical picture it is still in its infancy. The development of a qualitative chemically-sound understanding of a spin-phonon coupling is of fundamental importance in order to transfer the theoretical efforts of developing a quantitative first-principles theory to a synthetic strategy. Only very recently, Albino \textit{et al.}\cite{Albino2019} have provided a first-principles analysis of spin-phonon coupling in a series of V$^{4+}$ prototypical molecular qubits and discussed the role of molecular features such as coordination number and the chemistry of ligands. Additional theoretical insights in this direction have also been provided by Mirzoyan \textit{et al.}\cite{Mirzoyan2019} that have used Density Functional Theory to investigate the modulation of the ligand field by means of molecular distortions.

In order to further investigate the role of the ligands in the determination of spin-phonon coupling we here provide an in-depth analysis of the electronic structure of four molecular qubits with the same coordination shell symmetry but with different donor atoms, going from Oxygen to Sulphur, Selenium and Tellurium. After providing evidence for the reliability of the electronic structure method employed we proceed to analyse the contributions to the $g$-tensor and spin-phonon coupling coefficients across the series. In particular, we here interpret the mutli-reference electronic structure theory results with a simple ligand field model, providing an analytical relation between spin-phonon coupling coefficients and chemically-sound quantities such as static $g$-shifts, electronic ground-excited states energy splitting and chemical bond covalency. These results provide new fundamental insights into the nature of spin-phonon coupling, further demonstrating the potential of a combined theoretical and computational approach to the study of spin relaxation.   

\section*{Computational Methods}

The ORCA software~\cite{Neese2012} has been employed for all the calculations. The Complete Active Space Self Consistent Field (CAS-SCF) method and the N-Electron Valence Perturbation Theory (NEVPT2) have been employed as level of electronic structure theory, while Density Functional Theory with the PBE functional\cite{Perdew1996} has been used for the geometry optimizations and to produce the molecular orbitals initial guess for the CASSCF calculations. We have used the basis sets def2-TZVP for V, O, S, Se and Te, while the def2-SVP basis set have been used for C and H. A def2-TZVP/C auxiliary basis set was employed for all the elements\cite{Weigend2005}. The $\mathbf{g}$ tensor has been computed from effective Hamiltonian theory formalism. The results did not show any dependence on the introduction of larger basis sets and scalar relativistic effects. The spin-phonon coupling coefficients have been calculated as numerical derivatives of the $g$-tensor. Ten Cartesian displacements ranging from $\pm 0.01$ \AA. have been used to estimate the $(\partial \mathbf{g}/\partial X_{is})_{0}$ derivatives, where $X_{is}$ refers to the $s$ Cartesian component of the $i$-th atom and the subscript $0$ indicates that the derivatives are computed at the DFT optimized geometry. The $\mathbf{g}$ vs $X_{is}$ profiles have been fitted with a third order polynomial expression and the linear term had been retained. Coefficients with a regression error larger than 10\% were set to zero. Only the spin-phonon coupling of symmetry inequivalent atoms was computed explicitly.

\section*{Results}

\subsection*{Electronic structure and coordination bond in exa-coordinate V$^{4+}$ molecular qubits} 

\begin{figure}[h!]
\centering
\includegraphics[scale=1]{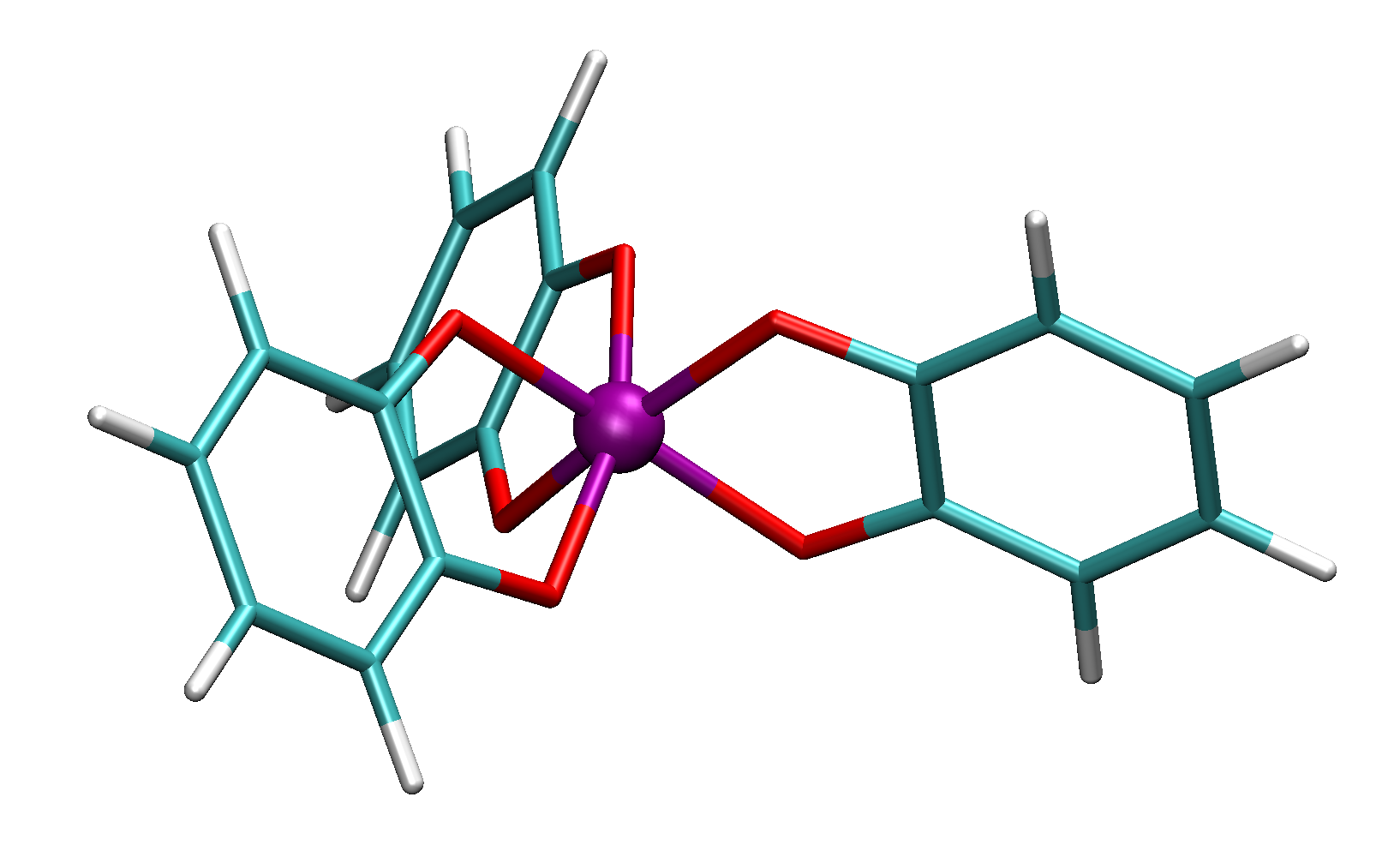}
\caption{\textbf{Optimized structure of V-O:} The Vanadium atom is coloured in purple, Oxygen atoms in red, Carbon atoms in green and Hydrogen atoms in white.}
\label{geo}
\end{figure}

In order to quantify the role of the first coordination shell in a systematic and quantitative computational way, we here consider four exa-coordinate V$^{4+}$ compounds. In all cases the central metal ion is coordinated by three ligands based on the di-anion catecholate, where the two RO$^{-}$ groups are replaced by RS$^{-}$, RSe$^{-}$ and RTe$^{-}$. Starting from the available X-ray crystallography structures, we optimise the geometry of the four complexes, hereafter indicated on the basis of the chemical bond involved (V-O, V-S, V-Se and V-Te). All the complexes share the same nominal $D_{3}$ symmetry, as reported in Fig. \ref{geo} for V-O, and show an increasing Metal-Ligand (ML) bond distance going from V-O to V-Te. Table \ref{bonds} shows the excellent agreement between calculated and experimental bond distances.

\begin{table} [!h]
	\small
	\centering
	\begin{tabular}{p{1.5cm}p{1.5cm}p{1.5cm}p{1.5cm}p{1.5cm}}	
		\hline
		& \textbf{V-O} & \textbf{V-S} & \textbf{V-Se} & \textbf{V-Te} \\
		\hline
		DFT & 1.97 \AA$ $ & 2.39 \AA$ $ & 2.51 \AA$ $ & 2.70 \AA$ $ \\
		X-ray & 1.95 \AA$ $ & 2.38 \AA$ $ & 2.49 \AA$ $ & -- \\
		\hline
	\end{tabular}		
	\caption{Calculated and experimental Metal-Ligand bond distances.}
	\label{bonds}
\end{table} 

The first step of our computational study consists in determining the most appropriate and accurate first-principles method to capture the fine details of both ground and excited electronic states. Complete Active Space Self Consistent Field (CASSCF) represents an extremely versatile framework where the variational electronic wave-function is expressed as series of Slater determinants built from a relatively small number of orbital ($n_{o}$) and electrons ($n_{e}$), \textit{i.e.} the active space, and indicated as CAS($n_{o}$,$n_{e}$). In the study of mono-nuclear coordination compounds, the active space is usually restricted to the $d$-like orbitals and electrons as a mean to capture static electronic correlation. Multi-reference perturbation theory, such as CASPT2 or NEVPT2, is then applied to the CASSCF wave-function in order to recover dynamical correlation. Although this represents the mainstream approach to the computation of spin Hamiltonian parameters of mono-nuclear complexes of transition metal and lanthanide ions, a minimal-CASSCF+NEVPT2 is not guarantee to capture all the relevant correlation needed to describe the complex metal-ligand bond. Here we attempt to obtain a converged description of the electronic structure of our class of compounds by progressively increasing the size of the active space.

\begin{figure}[h!]
\centering
\includegraphics[scale=1]{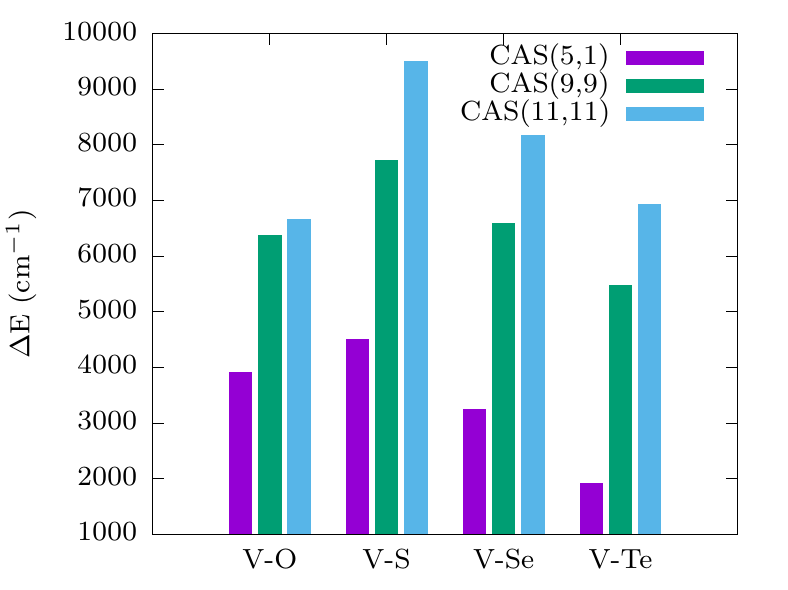}
\includegraphics[scale=1]{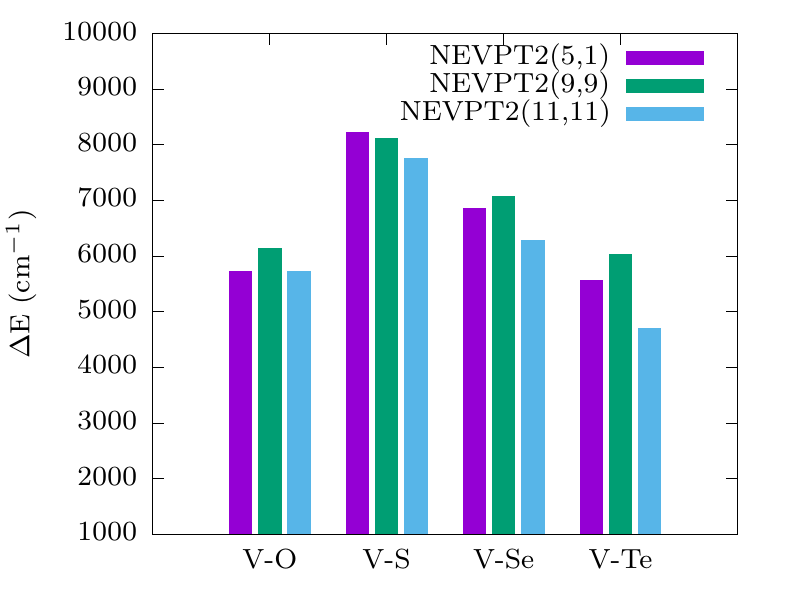}
\caption{\textbf{First electronic excited state energy splitting.}}
\label{E1}
\end{figure}

From a Ligand Field theory perspective, the use of a minimal active space composed by valence $d$-like orbitals and electrons corresponds to introducing non-bonding and anti-bonding orbitals between the metal ion's atomic $d$ orbitals and the ligands' $p$ orbitals. For V$^{4+}$ ions, possessing a $d^{1}$ electronic configuration, this leads to a (5,1) minimal active space. However, as also reported recently by Singh \textit{et al.}\cite{Singh2018}, the inclusion of ligand-like bonding orbitals into the active space generally provides a more balanced description of the covalent component of the coordination bond. The natural extension of the minimal active space then includes four bonding molecular orbitals that originates from the overlap of the ligands' fully occupied $p$-like orbitals and the empty $d$-like V$^{4+}$ orbitals, leading to a CAS(9,9). Finally, we also consider a CAS(11,11) where a doubly occupied bonding $d_{z^{2}}$-like orbital and an empty $d_{z^{2}}$ are introduced. The latter, belonging to a 4$d$ sets or orbitals instead of $3d$ orbitals, has the role of providing more radial flexibility to the wave-function describing the ground state. For all these active spaces we also perform a NEVPT2 calculation. In all cases we perform a state-average CASSCF calculation determining five solutions with spin multiplicity of two, corresponding to all the $d-d$-like ligand-field transitions. Fig. \ref{E1} reports the energy splitting of the first electronic excited state resulting for these calculations. The top panel, reporting the results for the CASSCF simulations before the NEVPT2 energy correction is applied, shows that the large effect of introducing bonding orbitals with large ligand character in the active space. Moreover, the size of this correction increases dramatically across the series moving from V-O to V-Te. The bottom panel of Fig. \ref{E1} shows instead the results after the NEVPT2 correction is applied. NEVPT2 does an overall good jobs at recovering correlation and the difference of energy splitting resulting from the use of different CASSCF reference wave-functions is significantly reduced. Interestingly the all the methods agree on the trend of the first excited state energy splitting, placing the one resulting  from the V-S as the larger one.  

We note that the empty $4d_{z^{2}}$ orbital included to CAS(11,11), although becoming increasingly important across the series (\textit{vide infra}), marginally contributes to the solutions under investigation and, therefore, we did not pursue further extension of the active space by including a full second $d$-shell. 

\begin{figure}[h!]
\centering
\includegraphics[scale=1]{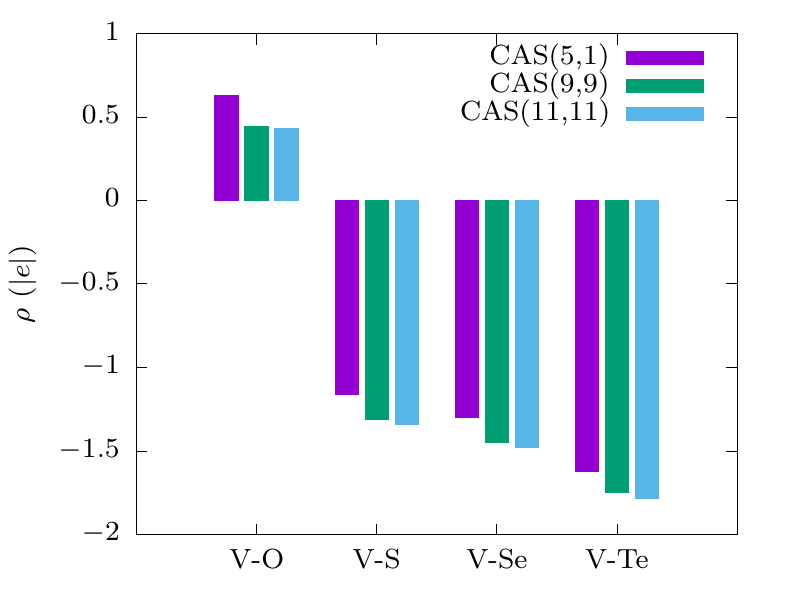}
\includegraphics[scale=1]{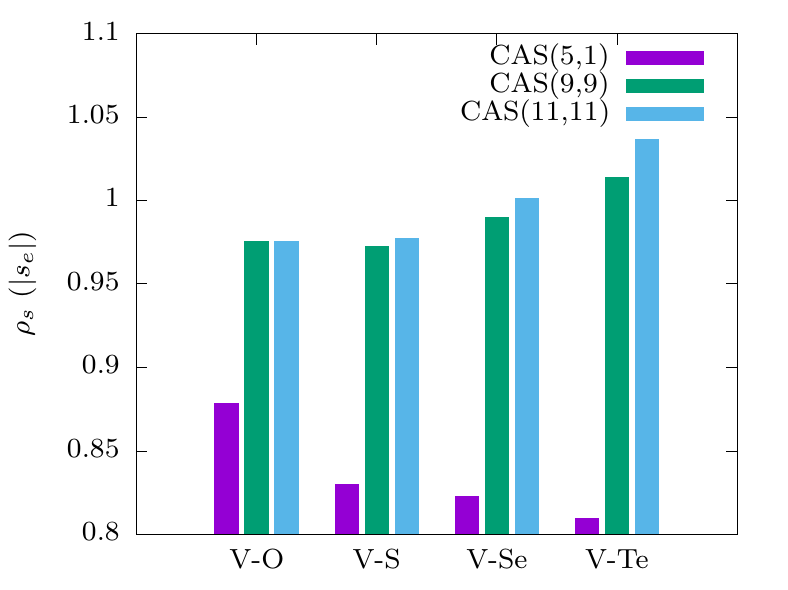}
\caption{\textbf{Vanadium charge and spin density:} The Vanadium local charge and spin densities computed with a Loewdin population analysis of the wave-function of state-average CASSCF calculations.}
\label{LOEW}
\end{figure}

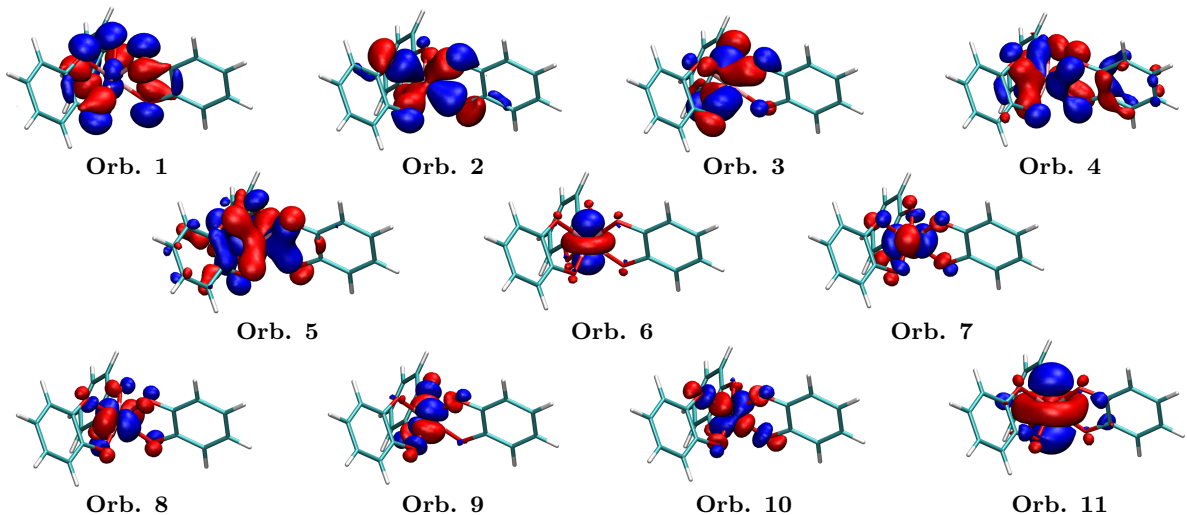
\begin{figure*}
\centering
\begin{tikzpicture}
\node at (0,0) {\includegraphics[scale=1]{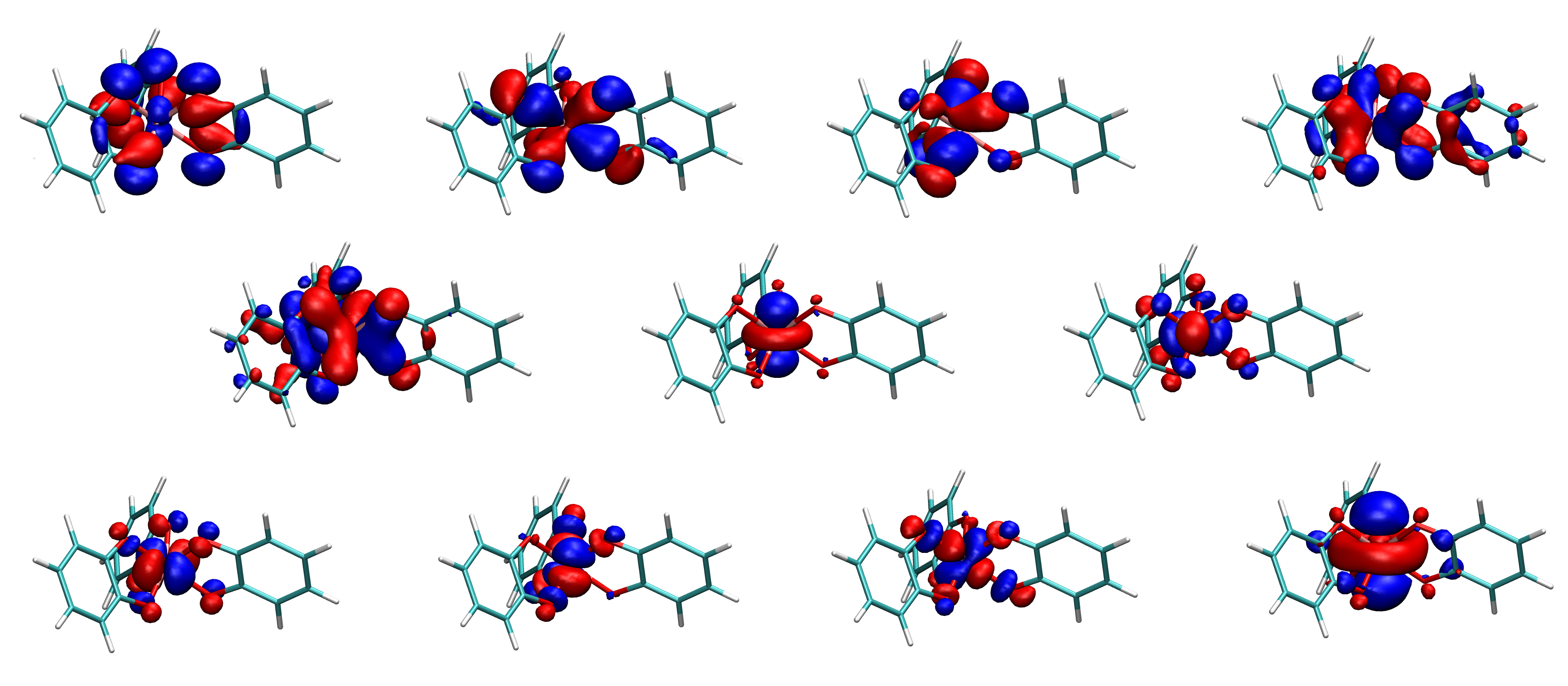}};
\node (A0) at (-6.2,1.1) {\textbf{Orb. 1}};
\node (A0) at (-2,1.1) {\textbf{Orb. 2}};
\node (A0) at (2,1.1) {\textbf{Orb. 3}};
\node (A0) at (6.2,1.1) {\textbf{Orb. 4}};
\node (A0) at (-4.2,-1.1) {\textbf{Orb. 5}};
\node (A0) at (0.25,-1.1) {\textbf{Orb. 6}};
\node (A0) at (4.5,-1.1) {\textbf{Orb. 7}};
\node (A0) at (-6.2,-3.4) {\textbf{Orb. 8}};
\node (A0) at (-2,-3.4) {\textbf{Orb. 9}};
\node (A0) at (2,-3.4) {\textbf{Orb. 10}};
\node (A0) at (6.2,-3.4) {\textbf{Orb. 11}};
\end{tikzpicture}
\caption{\textbf{V-O Active Natural Orbitals.}}
\label{MOS}
\end{figure*}
\begin{table*}
	\small
	\centering
	\begin{tabular}{p{1.cm}p{1.3cm}p{1.3cm}p{1.3cm}p{1.3cm}p{1.3cm}p{1.3cm}p{1.3cm}p{1.3cm}p{1.3cm}p{1.3cm}p{1.3cm}}	
		\hline
		& \textbf{Orb. 1} & \textbf{Orb. 2} & \textbf{Orb. 3} & \textbf{Orb. 4} & \textbf{Orb. 5} & \textbf{Orb. 6} & \textbf{Orb. 7} & \textbf{Orb. 8}
		& \textbf{Orb. 9} & \textbf{Orb. 10} & \textbf{Orb. 11} \\
		\hline
                \textbf{V-O} & 1.997 & 1.973 & 1.973 & 1.952 & 1.952 & 0.995 & 0.049 & 0.049 & 0.027 & 0.027 & 0.006 \\
		\textbf{V-S} & 1.994 & 1.942 & 1.941 & 1.936 & 1.935 & 0.987 & 0.067 & 0.067 & 0.059 & 0.058 & 0.015 \\
		\textbf{V-Se} & 1.993 & 1.934 & 1.934 & 1.926 & 1.926 & 0.986 & 0.075 & 0.075 & 0.067 & 0.067 & 0.016 \\
	        \textbf{V-Te} & 1.991 & 1.930 & 1.929 & 1.909 & 1.908 & 0.983 & 0.092 & 0.092 & 0.074 & 0.073 & 0.020 \\
		\hline
	\end{tabular}		
	\caption{Active natural orbital occupation number.}
	\label{nocc}
\end{table*}

Although the use of NEVPT2 nicely converges the size of the excited states' energy splitting, it does not improve the wave-function. Looking forward to the calculation of magnetic properties this represents a limitation and a careful analysis of the orbitals must also be carried out. The top and bottom panels of Fig. \ref{LOEW} reports the charge and spin population, respectively, for the Vanadium atom across the series as obtained from the Loewdin orbital analysis. These results show that CAS(9,9) and CAS(11,11) provide a well converged picture of the molecular charge and spin density, while the minimal CAS(5,1) capture a qualitatively different spin density trend. According to this analysis we can conclude that the description of the electronic structure of this family of molecules by CAS(11,11)+NEVPT2 is well converged and reliable. 

\begin{figure}[h!]
\centering
\includegraphics[scale=1]{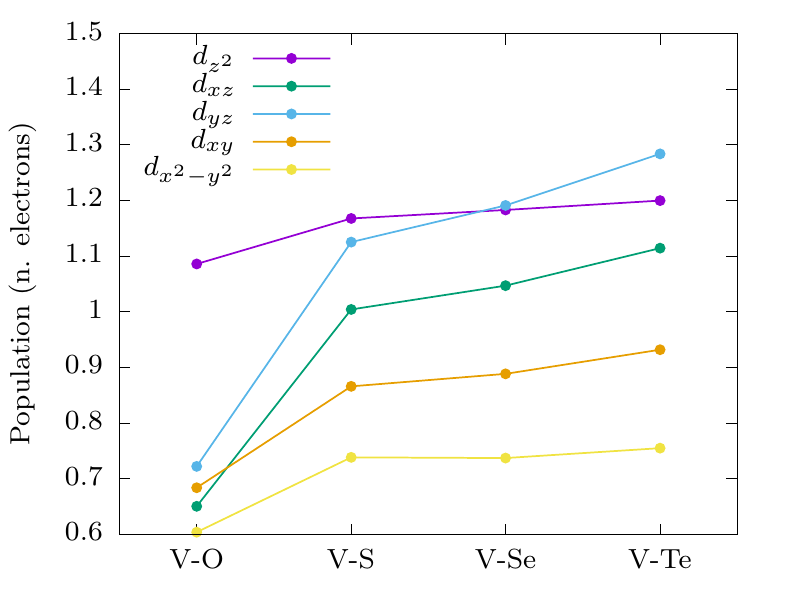}
\caption{\textbf{Vanadium $d$ orbitals population:} The local occupation of the $d$ orbitals computed with a Loewdin population analysis of the wave-function of ground-state state-selected CAS(11,11) calculation.}
\label{dorbs}
\end{figure}

In order to address the nature of the coordination bond we next perform ground state state-selected calculations with CAS(11,11)+NEVPT2. Fig. \ref{dorbs} reports the active natural orbitals for V-O. As expected from Ligand Field theory arguments, it is possible to recognize five bonding orbitals (Orbs. 1-5), five anti-bonding orbitals (Orbs. 6-10) and a single $4d_{z^{2}}$ (Orb. 11). Although quantitative differences between the active orbitals of different compounds exist, the natural orbitals reported in Fig. \ref{dorbs} are a good qualitative representation for the entire series. Table \ref{nocc} reports the occupation number of each active orbital for the entire series and shows an increasing tendency to occupy anti-bonding-like orbitals with respect to bonding ones, going from V-O to V-Te. This result can be interpreted as an increasing level of Ligand to Metal Charge Transfer (LMCT) character of the ground state wave-function across the series. 
Fig. \ref{dorbs} reports the composition of the ground state wave-function in terms of $d$-like orbitals, as provided by the Loewdin analysis. The variation of the atomic $d$-like orbitals' population across the series of compounds is in close agreement with the analysis of the state-average solution and reveals that the population of $d$-like orbitals in the ground state wave-function progressively increases. 

Both this trend and the increasing LMCT character of the ground state can be interpreted as a progressively increase in the covalent character of the coordination bond. In fact, the ion V$^{4+}$ has a $d^{1}$ configuration and principally contribute to the chemical bond through empty $d$ orbitals, which accept electron density coming from the ligands' orbitals.

\subsection*{$G$-tensor, spin-phonon coupling and ligand-field analysis} 

The four V$^{4+}$ complexes here investigated all have a spin-1/2 ground state and, as such, their single-ion magnetic properties can be described by a spin Hamiltonian including the Zeeman and Hyperfine interactions. However, in this work we will limit ourselves to the study the high external field limit, where only the Zeeman contribution is expected to play a dominant role in the spin relaxation\cite{Lunghi2019b}. To this end, we here report a comprehensive study of the $g$-tensor and the spin-phonon coupling originated from its modulation by the atomic displacements.

Table \ref{gtens} reports experimental available $g$-tensor principal components, extracted from cw-EPR experiments\cite{Atzori2018a,Fataftah2019a}, together with the values computed with CAS(11,11)+NEVPT2. The qualitative trend for V-O, V-S and V-Se is reproduced, although a significant overestimation of the calculated $g$-shift ($\delta g$), \textit{i.e.} the distance of the $g$-tensor's eigenvalues from the free electron $g_{e}$ value, in V-O is observed. 

Spin-phonon coupling coefficients are defined as the derivatives of the spin Hamiltonian terms with respect to the atomic displacements and quantify the interaction between atomic displacements and spin degrees of freedom. Although the former are usually described in terms of normal modes of vibration, we here perform a differentiation in terms of Cartesian coordinates $X_{iv}$, where the coefficients $i$ and $v$ run on the atoms and the Cartesian coordinates, respectively. The use of Cartesian coordinates instead of normal modes makes it possible to focus on the sole contribution coming from the geometrical distortions and obtain a transparent comparison of spin-phonon coupling across the series. In particular this analysis eliminates the complexity introduced by the differences in energy and nature of the normal modes of different molecules. According to this definition the spin-phonon coupling coefficients are simply $(\partial g_{st}/\partial X_{iv})_{0}$, where the indexes $s$ and $t$ run on the three Cartesian coordinates of the $g$-tensor and the subscript $0$ refers to the derivatives taken at the equilibrium geometry. Following previous analysis we also introduce a molecular-average spin-phonon coupling coefficient as:

\begin{equation}
||\partial g||=\sum_{s,t}\sum_{i,v}\Big(\frac{\partial g_{st}}{\partial X_{iv}}\Big)^{2}_{0}\:.
\end{equation}

The average spin-phonon coupling coefficients, calculated for all the molecules in the series, are reported in Table \ref{gtens} and show a nice correlation with the size of $\delta g$. This correlation was already reported by Albino \textit{et al.}\cite{Albino2019} for V-O and other V$^{4+}$ complexes and it is here confirmed by the study of the three new molecules V-S, V-Se and V-Te.

\begin{table} [!h]
	\small
	\centering
	\begin{tabular}{p{1.9cm}p{1.5cm}p{1.5cm}p{1.5cm}p{1.5cm}}	
		\hline
		& \textbf{V-O} & \textbf{V-S} & \textbf{V-Se} & \textbf{V-Te} \\
		\hline
		&\multicolumn{4}{c}{Experimental Parameters} \\		
		$g_{xx}$ & 1.945(1) & 1.9698 & 1.950 & -- \\	
		$g_{yy}$ & 1.945(1) & 1.9698 & 1.960 & -- \\			
		$g_{zz}$ & 1.989(2) & 1.9878 & 1.955 & -- \\
		\hline
		&\multicolumn{4}{c}{Calculated Parameters}\\		 	
		$g_{xx}$ & 1.906 & 1.953 & 1.942 & 1.951 \\
		$g_{yy}$ & 1.907 & 1.953 & 1.942 & 1.951 \\
		$g_{zz}$ & 1.997 & 1.999 & 1.997 & 1.998 \\
		\hline
		&\multicolumn{4}{c}{Spin-phonon Coupling} \\		 	
		$||\partial g||$ (\AA$^{-2}$) & 0.219 & 0.034 & 0.135 & 0.052 \\
		\hline
	\end{tabular}		
	\caption{Calculated and experimental $g$-tensor principal components and averaged spin-phonon coupling}
	\label{gtens}
\end{table} 

In order to provide a chemical interpretation of the results relative to the $g$-tensor we analyse them in terms of the Ligand-Field theory. In this framework and according to second-order perturbation theory, the components of the $g$-tensor assume a different value from the free electron $g$ factor $g_{e}$ due to the presence of spin-orbit coupling and low-lying electronic excited states with the same spin multiplicity as the ground state\cite{Neese1998}. Eq. \ref{gxx} shows the contribution to the $g_{xx}$ component of the $g$-tensor due to the presence of a single excited state. 

\begin{equation}
 g_{xx}=g_{e}-\frac{\zeta}{S\Delta}\langle \psi_{0} | \hat{l}_{x} | \psi_{1} \rangle\langle \psi_{1} | \hat{l}_{x} | \psi_{0} \rangle\:,
\label{gxx}
\end{equation}

where $\hat{l}_{x}$ is the electronic angular momentum with respect to the Vanadium nucleus, $\Delta$ is the energy separation between the ground state and the excited state characterized by two wave-functions $\psi_{0}$ and $\psi_{1}$, respectively. Finally, $\zeta$ is the free-ion spin-orbit coupling constant and S is the value of the molecular ground-state spin.

In the Ligand-Field theory the ground-state and excited state wave-functions $\psi_{0}$ and $\psi_{1}$ can be written as

\begin{align}
& | \psi_{0} \rangle = \alpha |d_{z^{2}} \rangle + \alpha' |\phi_{0} \rangle \\
& | \psi_{1} \rangle = \beta |d_{yz} \rangle + \beta' |\phi_{1} \rangle
\label{ligwave}
\end{align}

where $\alpha$ and $\beta$ describe the contribution of V$^{4+}$'s atomic orbitals that participate to $g_{xx}$, while $\alpha'$ and $\beta'$ represent the contribution of the ligands' orbitals. Assuming that only the portion of the orbitals localized on the V$^{4+}$ centre can contribute to the integrals of Eq. \ref{gxx} and substituting Eq. \ref{ligwave} into Eq. \ref{gxx}, one obtains that 

\begin{equation}
\delta g_{xx}=g_{xx}-g_{e}=-6\zeta\frac{\alpha^{2}\beta^{2}}{\Delta}=-\zeta_{eff}\Delta^{-1}\:, 
\label{covshift}
\end{equation}

where the coefficients $\alpha^{2}$ and $\beta^{2}$ accounts for the delocalization of molecular orbitals on the ligands and the consequent reduction of the effective spin-orbit coupling constant $\zeta_{eff}$. This is known as the relativistic Nephelauxetic effect\cite{Neese1998}. 

\begin{figure}[h!]
\centering
\includegraphics[scale=1]{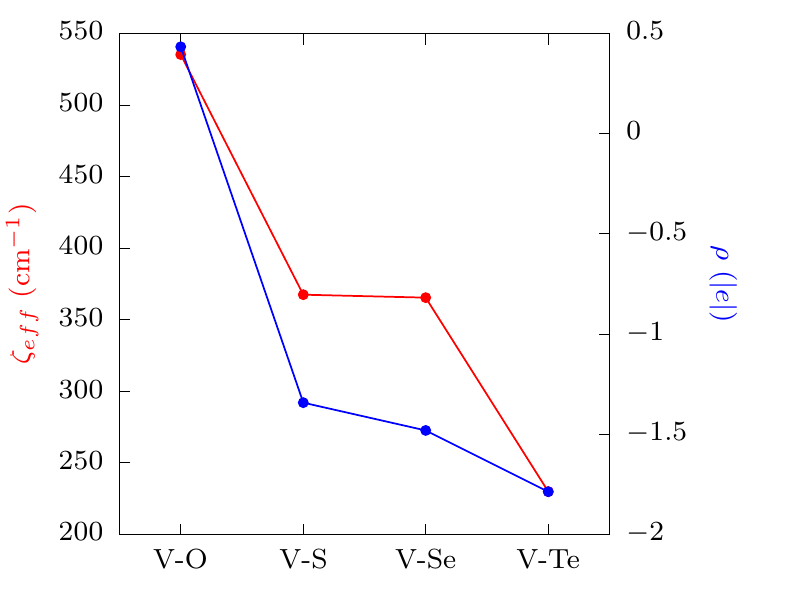}
\caption{\textbf{Correlation between covalency and V electron density:} The left y-axis reports the Vanadium effective spin-orbit coupling constant across the series, while the right y-axis reports the local charge density of Vanadium as extracted from a Loewdin population analysis for the state-average CASSCF(11,11) calculation.}
\label{alpha}
\end{figure}

Combining Eq. \ref{covshift} with the numerical results of Table \ref{gtens} and the calculated energy splitting reported in Fig.\ref{E1} we obtain an estimation of $\zeta_{eff}$, accounting for the changes in the covalency term $\alpha^{2}\beta^{2}$ across the series. $\zeta_{eff}$ is found to decrease across the series in a very similar fashion as to the population of the d-like orbitals. The two figures are plotted together in Fig. \ref{alpha}. Despite the big approximation made in the model expressed by Eqs. \ref{gxx} and \ref{ligwave} this analysis show an overall agreement with the interpretation of covalency increasing along the series. 

\begin{figure}[h!]
\centering
\includegraphics[scale=1]{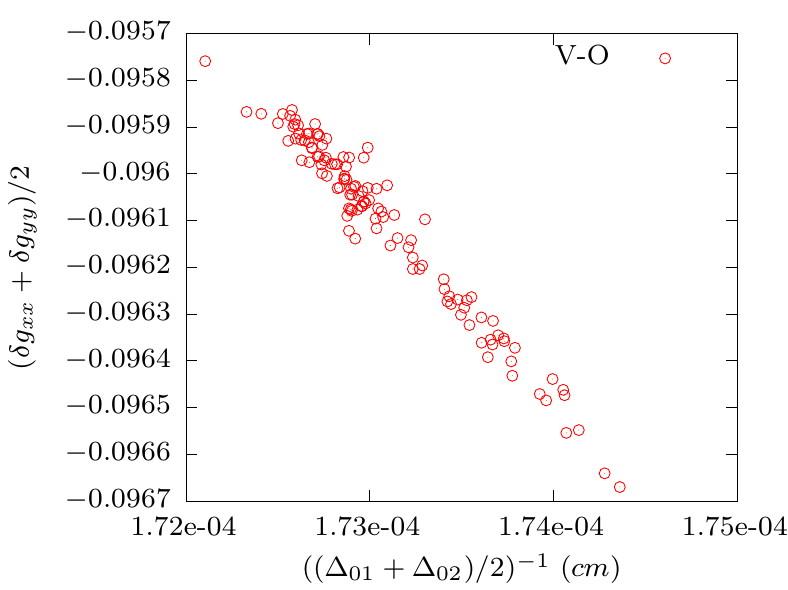}
\caption{\textbf{Correlation between $g$-shift and excited state energy:} The average of $\delta g_{xx}$ and $\delta g_{yy}$ is plotted against the inverse of the average energy of the first two quasi-degenerate electronic excited states for the V-O complex.}
\label{zetaeff}
\end{figure}

The same Ligand-Field model used to interpret the various contributions to the $g$-tensor can also shed light on the spin-phonon coupling coefficients and, in particular, on the origin of the correlation between the average spin-phonon coupling coefficients and the static $g$-shifts values. By definition the spin-phonon coupling coefficients are the derivatives of the $g$-tensor with respect to the atomic positions. As also very recently investigated by Mirzoyan \textit{et al.}\cite{Mirzoyan2019}, both the numerator and denominator of expression \ref{covshift} are potentially dependent on the atomic positions and, therefore, potentially contributing to the spin-phonon coupling. In order to investigate this possibility we plot the average value of $\delta g_{xx}$ and $\delta g_{yy}$ against the inverse of the average energy of the first two quasi-degenerate excited states, that presumably contribute to $\delta g$, for all the distorted geometries used for the numerical differentiation of the $g$-tensor. The results of this analysis for V-O,  as representative for the entire series, are reported in Fig. \ref{zetaeff} and they show a linear relation between the two quantities. This result illustrates the constant nature of the covalency term with respect to the atomic displacements, \textit{i.e.} $\alpha(X_{is})\sim \alpha_{0}$. Under this condition, the derivative of Eq. \ref{covshift} with respect to the  atomic displacements leads to

\begin{equation}
 \Big(\frac{\partial g_{xx}}{\partial X_{is}}\Big)_{0}=6\zeta \frac{\alpha^{2}\beta^{2}}{\Delta^{2}}\Big(\frac{\partial \Delta}{\partial X_{is}}\Big)_{0}=-\frac{\delta g_{xx}}{\Delta}\Big(\frac{\partial\Delta}{\partial X_{is}}\Big)_{0}\:.
\label{gder}
\end{equation}

From Eq. \ref{gder} it is now evident the dependence of the derivatives of the $g$-tensor on the magnitude of the $g$-shift themselves, in agreement with the numerical results. Eq. \ref{gder} also shows a quadratic dependence of the spin-phonon coupling coefficients on the spin energy splitting and a liner dependence on its gradient, once more highlighting the important of this quantity as a figure of merit for the spin relaxation process. Interestingly, extending Eq. \ref{gder} to the next order of derivation, the dependence on the static $g$-shift is preserved. This implies that the use of $g$-shift as an approximate proxy for spin-phonon coupling efficiency remains valid also when the direct mechanism is overcome by Raman relaxation.  

\section*{Discussion and Conclusions}  

The individuation of simple chemical guidelines to the extension of spin-lattice relaxation time is a fundamental step towards the design of new molecular qubits. From a theoretical perspective, the key role of spin-phonon coupling and molecular vibrations in the determination of spin-lattice relaxation time is now well understood and it is clear that they should be used a starting point for a quantitative discussion of spin dynamics. However, a clear connection between spin-phonon coupling and a simple chemical picture had yet to be provided. While multi-reference electronic structure theory provides a robust framework for the prediction of all the relevant parameters that drive spin relaxation, chemical synthesis invariably needs quantum mechanical theory to be translated into a more intuitive chemical language\cite{Kumar2017}. 

Eq. \ref{gder}, obtained from a Ligand-Field theory model and validated through multi-reference electronic structure theory, offers such a chemistry-wise perspective and clearly highlights the major contributions to the spin-phonon coupling in spin-$1/2$ originating from the atomic modulation of the $g$-tensor. Here, by showing that molecular covalency remains almost constant during molecular motion, we were able to demonstrate that spin-phonon coupling coefficients are proportional to the static $g$-shifts and overall depend on i) the squared inverse of the excited states energy splitting, ii) the derivatives of the excited states energy splitting with respect to atomic displacements and iii) the chemical bond level of covalency. This analysis, also reported in the work of Mirzoyan \textit{et al.}\cite{Mirzoyan2019}, further supports the findings of Albino \textit{et al.}\cite{Albino2019} and provides a more quantitative ground for study of spin-phonon coupling in terms of experimental observables such as the static $g$-shifts or electronic optical excitations. Interestingly, the correlation between the static $g$-shift and spin relaxation time was empirically observed in the context of organic radicals\cite{Schott2017}.

Eq. \ref{gder} thus suggests the selection of coordination compounds that favour large covalency and large energy splitting at the same time. The reduction of spin-phonon coupling through the control of the ground-excited states energy splitting derivatives is arguably more complex. However, the already proposed strategy of designing rigid molecular complexes is expected to act exactly in this respect by \textit{freezing} the molecular motions responsible for that contribution\cite{Lunghi2017a}. In this respect it is also important to remark that the contributions to spin relaxation discussed here neglect the role of phonon's thermal population. The low-energy vibrational density of state plays a key role in determining the overall spin relaxation behaviour and both the nature of spin-phonon coupling discussed here and the thermal population of intra-molecular vibrations need to be taken into account in order to quantitatively interpret spin-lattice relaxation\cite{Lunghi2019b}.

Our theoretical and computational analysis of the contributions to spin-phonon coupling also makes it possible to rationalize a body of recent literature that proposes a variety of apparently orthogonal strategies to improve spin-lattice relaxation time in magnetic molecules. For instance, in light of Eq. \ref{gder}, several experimental works reporting respectively $g$-tensor isotropy\cite{Ariciu2019}, large molecular covalency\cite{Fataftah2019a} and molecular rigidity\cite{Bader2014,Atzori2016b} as primary ingredients for long spin-lattice relaxation times in molecular qubits, can be reconciled by recognizing that they all point to different contributions to spin-phonon coupling, \textit{i.e.} the fundamental physical quantity that drives spin relaxation.

In conclusion, we here investigated a series of four V$^{4+}$ compounds with a single unpaired electron in the d-shell that represent potential building blocks for molecular-based quantum architectures. We established a robust description of the coordination bond and electronic excited states by means of multi-reference electronic structure theory and were able to identify ligand-field-like contributions to the $g$-tensor and spin-phonon coupling coefficients. Most importantly, we formally demonstrated the correlation between $g$-shift values and spin-phonon coupling coefficients. This combined Ligand-Field and multi-reference analysis of the results provided a chemical interpretation of the spin-phonon coupling coefficients revealing the energy separation between the ground and excited electronic states and bond covalency as two fundamental figure-of-merit for the optimization of spin-lattice relaxation time in molecular qubits.

\noindent
\textbf{Data Availability}\\
All the relevant data discussed in the present paper are available from the authors upon request. \\

\noindent
\textbf{Acknowledgements}\\
This work has been sponsored by AMBER (grant 12/RC/2278\_P2). Computational resources were provided by the Trinity Centre for High Performance Computing (TCHPC) and the Irish Centre for High-End Computing (ICHEC). We also acknowledge the MOLSPIN COST action CA15128.\\

\noindent
\textbf{Conflict of interests}\\
The authors declare that they have no competing interests.


\end{document}